\documentclass[fleqn,twoside]{article}
\usepackage{espcrc2}
\usepackage{epsfig}

\newcommand{\AmS}{{\protect\the\textfont2
  A\kern-.1667em\lower.5ex\hbox{M}\kern-.125emS}}

\newcommand{\etal} {{\it et al.}}

\title{Cosmic Rays from the Knee to the Ankle -- Status and 
Prospects}

\author{Karl-Heinz Kampert\address[NUW]{Department of Physics, 
        Bergische Universit\"at Wuppertal \\ 
        D-42119 Wuppertal, Germany}%
        \thanks{email: {\tt kampert@uni-wuppertal.de}}}
       
\begin{document}

\begin{abstract}
Recent progress in cosmic ray physics covering the energy range
from about $10^{14}$\,eV to $10^{19}$\,eV is reviewed.  The most
prominent features of the energy spectrum are the so called
`knee' at $E \simeq 3 \cdot 10^{15}$\,eV and the `ankle' at few
$10^{18}$\,eV. Generally, the origin of the knee is understood as
marking the limiting energy of galactic accelerators and/or the
onset of increasing outflow of particles from the galaxy while
the ankle is considered to mark the transition from galactic to
extragalactic cosmic rays.  Alternative theories do exist and
shall be sketched.  A key observable to answer the still open
questions about the cosmic ray origin and to discriminate between
various models is given by measuring the chemical composition or
-- more directly -- by measuring energy spectra of individual
cosmic ray mass groups.  The status of present analyses is
critically discussed and new experimental endeavors carried out
in order to improve both the statistics and the quality of data
particularly at energies above the knee will be summarized.
\vspace{1pc}
\end{abstract}

\maketitle

\section{INTRODUCTION}

The cosmic ray (CR) energy spectrum extends from a few hundreds
MeV to above $10^{20}$~eV. Over this wide range of energies the
intensity drops by more than 30 orders of magnitude.  Despite the
enormous dynamic range covered, the spectrum appears rather
structureless and can be well approximated by broken power-laws
$dN/dE \propto E^{-\gamma}$.  Up to energies of a few
$10^{14}$~eV the flux of particles is sufficiently high to enable
measurements of their elemental distributions by high flying
balloon- or satellite-borne experiments.  Such studies have
provided important information about the origin and transport
properties of CRs in the interstellar medium.  Two prominent
examples are ratios of secondary to primary elements, such as the
B/C-ratio, which are used to extract the average amount of matter
CR-particles have traversed from their sources to the solar
system, and are relative abundances of radioactive isotopes, such
as\ $^{10}$Be to stable $^{9}$Be or $^{26}$Al to stable
$^{27}$Al, which carry information about the average `age' of
CRs.  With many new complex experiments taking data or starting
up in the near future and with a possibly new generation of long
flying balloons, this remains a vital field of research.

\begin{figure*}[t]
\centering
\includegraphics[width=120mm]{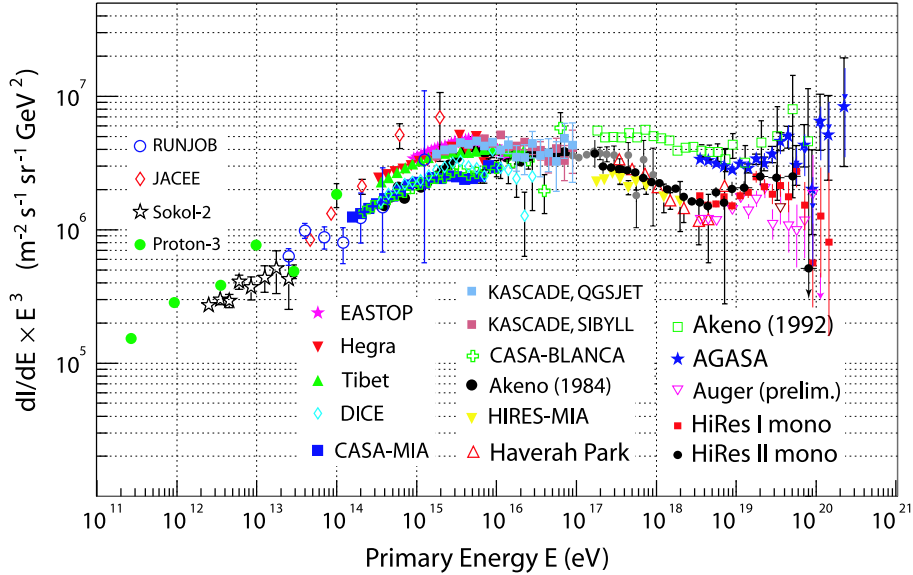}
\vspace*{-7mm}
\caption[xx]{The all-particle CR energy spectrum weighted by
$E^{3}$ showing the knee at $3\cdot10^{15}$~eV, a possible second
knee at $\sim 10^{17}$~eV, the ankle at about 
$3\cdot10^{18}$~eV, and the GZK-region near $6 \cdot 
10^{19}$\,eV. References are given in the text.}
\label{fig:all-particle}
\end{figure*}

Above a few times $10^{15}$~eV the flux drops to only one
particle per square metre per year.  This excludes any type of
`direct observation' even in the near future, at least if high
statistics is required.  On the other hand, this energy is large
enough so that secondary particle cascades produced in the
atmosphere penetrate with a footprint large enough to be detected
by an array of detectors on the ground.  Such an extensive air
shower (EAS) array typically has dimensions of a fraction of a
square kilometre to more than 1000 square kilometres and can be
operated for many years to detect fluxes down to 1 particle 
per square kilometre per century or less.

The most prominent features of the CR energy spectrum fall into
the energy range covered by EAS experiments.  The steepening of
the slope from $\gamma \cong 2.7$ to $\gamma \cong 3.1$ at about
$3\cdot 10^{15}$~eV is known as the `knee'.  It was first deduced
from observations of the shower size spectrum made by Kulikov and
Khristianson {\etal} in 1956 \cite{kulikov56} but it still
remains unclear as to what is the cause of this spectral
steepening and even as to what are the sources of the high energy
CRs at all.  At an energy above $10^{18}$~eV the spectrum
flattens again at what is called the `ankle'.  Because of the
large size and/or magnetic field required to accelerate and
confine charged particles above $10^{18}$\,eV, the origin of CRs
above the ankle is generally considered to be of extragalactic
(EG) nature.  Finally, the question whether the spectrum extends
beyond the Greisen-Zatsepin-Kuzmin threshold of $6\cdot
10^{19}$~eV \cite{GZK} is currently among the foremost questions
in astro-particle physics as is reflected also by the number of
presentations given at this conference.

The main purpose of this paper is to review the experimental data
in the energy range below the GZK-threshold, i.e.\ from about
$10^{14}$ to $10^{19}$\,eV. We shall discuss the energy spectrum,
chemical composition, and anisotropies in their arrival
directions and critically examine the astrophysical implications
by taking into account the systematical uncertainties of the
data.

\section{THE KNEE REGION}

Mainly for reasons of the required power the dominant
acceleration sites of CRs are generally believed to be shocks
associated with supernova remnants (SNR).  Naturally, this leads
to a power law spectrum as is observed experimentally.  Detailed
examination suggests that this process is limited to $E_{0}/Z
\sim 10^{14}$\,eV \cite{Lagage-83,Berezhko-00} for standard
galactic SNRs.  This value can be extended upward with a number
of mechanisms, for example by introducing higher magnetic fields,
larger sources, quasi-perpendicular shocks, reacceleration by
multiple sources, etc.  However, these assumptions and their
effects are not free of debate and possibly, something more
fundamental may be incorrect with the suggested supernova (SN)
picture and its shock value $E_{0}$.  In any case, if there is a
typical maximum energy which depends linearly on $Z$\/ for
reasons of magnetic confinement, then the spectrum of CR nuclei
must become heavier with increasing energy as the hydrogen cuts
off first and then increasingly heavier nuclei reach their
acceleration (or confinement) limits.

A change in the CR propagation with decreasing galactic
containment at higher energies has also been considered.  This
increasing leakage results in a steepening of the CR energy
spectrum and again would lead to a similar scaling with the
rigidity of particles, but would in addition predict anisotropies
in the arrival directions of CRs with respect to the galactic
plane.

Besides such kind of `conventional' source and propagation models
\cite{drury94b,berezhko99} several other hypotheses have been
discussed in the recent literature.  These include the
astrophysically motivated single source model of Erlykin and
Wolfendale \cite{erlykin97a} trying to explain possible
structures around the knee by a single recent and nearby SN, as
well as several particle physics motivated scenarios trying to
explain the knee due to different kinds of CR-interactions, e.g.\
by photodisintegration at the source \cite{candia-02} or by
sudden changes in the character of high-energy hadronic
interactions during the development of EAS \cite{nikolsky95}.

Recently, the `Cannonball' model of CRs has been suggested as a
radically different theory of CR origin \cite{Dar-06}.  It is
inspired by mounting observational evidence that, in addition to
the ejection of a non-relativistic spherical shell, the explosion
of core-collapse SNae results in the emission of highly
relativistic bipolar jets of plasmoids of ordinary matter, the
`Cannonballs' (CB).  As the CB with a typical half of the Mercury
mass propagates at relativistic speed through the interstellar
medium, it encounters electrons, protons, and nuclei kicking them
up to higher energies elastically by magnetic deflection.  These
newly born CRs are then subject to propagation effects, similarly
as in `classical' theories.  It is argued that this very simple
concept explains all observed properties of non-solar CRs at all
observed energies.  There are two important differences to the
conventional models: a) because of the specific kinematics of
particle acceleration, the maximum energy of CRs (and thereby the
knee positions) scale with the mass $A$ of CRs rather than with
their charge $Z$, b) since the CBs propagate rapidly from the
inner SN and GRB realm of the Galaxy into its halo or beyond
converting ISM particles to high energy CRs all along their
trajectories, there is a much lower level of CR-anisotropy
expected than in the traditional SN picture of CRs. 

Indeed, the low level of CR anisotropy even at energies above the
knee is a long standing problem \cite{Hillas-05}.  Generally, the
observed spectrum $\phi(E)$ and the source spectrum $Q(E)$ are
considered to be connected by a relation of the form $\phi(E) =
Q(E) \times \tau_{\rm esc}(E)$.  A simple power-law fit of the
escape time to the available data gives $\tau_{\rm esc}(E)
\propto E^{-\delta}$ with $\delta \approx 0.6$.  Extrapolating
$\tau_{\rm esc}$ to $10^{15}$\,eV, for example, would lead to a
value almost as small as the light travel time across the
galactic disk, implying a much larger anisotropy than is
observed.

From the discussion above it is obvious, that an answer to the
question about the origin of the knee is of key importance to
reveal information about the origin of galactic cosmic rays in
general.  Experimental access to such questions is provided by
measurements of charged cosmic rays (the classical nucleonic
component) and $\gamma$-rays by experiments above the atmosphere,
and by the observation of air showers initiated by high-energy
particles in the atmosphere.

A wealth of information on potential cosmic-ray sources is
provided by recent measurements of TeV $\gamma$-rays employing
imaging atmospheric Cherenkov telescopes, most notably from the
H.E.S.S.\ experiment.  Their observation of the morphologies and
energy spectra of the shell type SNRs RX J1713.723946
\cite{Aharonian-04,Aharonian-06} and RX J0852.0-4622
\cite{Aharonian-05a} are well in agreement with the idea of
particle acceleration in the shock front.  The spectra extend up
to energies of 10~TeV and provide evidence for the existence of
particles with energies beyond 100~TeV at the shock front that
emerged from the supernova explosions.  However, an unequivocally
proof for acceleration of hadrons is still missing and questions
arise also about the low number of established SNRs showing TeV
$\gamma$-ray emission.  For example, a recent Galactic plane
survey of H.E.S.S.\ \cite{Aharonian-05b} reveals no SNRs brighter
than these two in the region covered.  This apparent deficit of
TeV-bright SNRs may pose some problems in explaining the high
energy budget of galactic CRs.  Remember that about 10\,\% of the
mechanical energy released by the population of Galactic
supernovae needs to be converted into CRs if {\em all\/} SNRs are
sites of CR acceleration.  Any reduction in the number of
TeV-bright SNRs needs to be compensated for by a corresponding
factor in the already large value of the CR acceleration
efficiency.

To undoubtedly establish SNRs as the sites of CR acceleration and
in order to constrain the conventional SN acceleration model from
other proposed mechanisms, precise measurements of the primary CR
energy spectrum and particularly of the mass composition as a
function of energy are needed.  Significant progress has been
made here as well in recent years, but the situation is far from
being clear.

\subsection{Comparison of direct and indirect measurements}

Cosmic ray measurements on balloons and spacecraft have an
important advantage over ground-based air shower experiments:
They detect the {\em primary} CR particles and measure its charge
{\em directly}.  This is because spacecraft experiments perform
the measurement above the atmosphere and balloon-borne
experiments typically perform their measurements with residual
atmospheres of only $\sim 5$-10\,g/cm$^{2}$.  This is a
relatively small value compared to the typical hadronic
interaction length of $\lambda_{I} \sim 90$\,g/cm$^{2}$ so that
corrections for interactions above the instruments are of minor
importance, at least for light particles, such as protons and He
nuclei.  This advantage is paid for at the expense of lacking
statistics at high energies.  For example, the largest of the
current generation of balloon-borne detectors, TRACER
\cite{Mueller-05}, reaches a sensitive volume of $2 \times 2
\times 1.2$\,m$^{3}$.  It has been flown successfully for 14 days
exposure from the Antarctic in 2003 and from Sweden in summer
2006.  The first 14 days flight time resulted in an exposure of
$\sim 75 \rm{\,m}^{2} \rm{~sr~days}$ and allowed to measure e.g.\
oxygen nuclei up to $\sim 320$\,TeV and iron nuclei up to $\sim
70$\,TeV.

\begin{figure}[t]
\centerline{\epsfxsize=\columnwidth\epsfbox{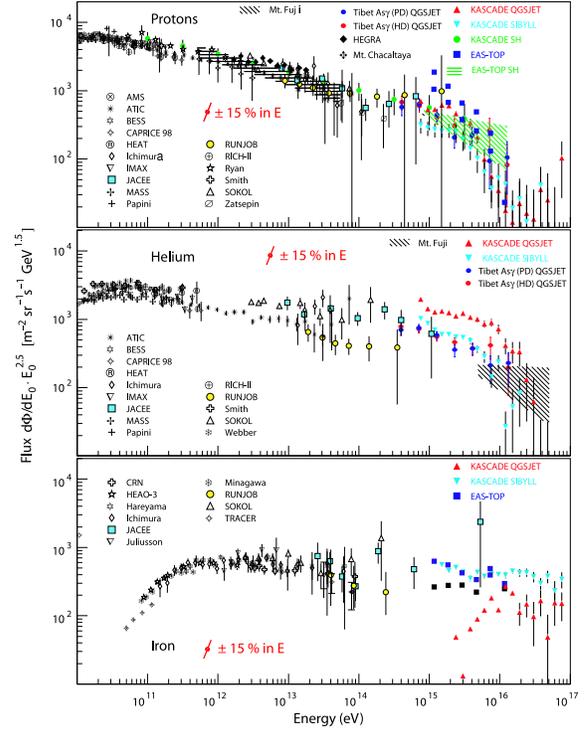}}
\vspace*{-7mm} \caption[xx]{Proton (top), helium (centre), and
iron (bottom) spectra from direct experiments compared to EAS
data (based and updated from \cite{Hoerandel-05}).  The single
diagonal error bar in each panel indicates the effect of a $\pm
15$\,\% uncertainty in the energy scale.
\label{fig:p-he-fe}}
\end{figure}

The largest exposure of all direct experiments has been reached
by the Japanese American Cooperative Emulsion Experiment JACEE
\cite{Asakimori-98} and the RUssian-Nippon JOint Balloon
collaboration RUNJOB \cite{Derbina-05}.  JACEE flew a series of
thin ($\sim 8.5$ radiation lengths) emulsion/X-ray film
calorimeters on 15 flights during 1979-1994 and has reached an
exposure of $\sim 664 \rm{\,m}^{2} \rm{~hrs}$ from 11 analysed
flights.  Taking the zenith angle acceptance out to $\tan \theta
\sim 72{\rm -}79^{\circ}$ into account, this relates to approx.\
$\sim 80 \rm{\,m}^{2} \rm{~sr~days}$.  RUNJOB flew roughly a
similar set of X-ray films and emulsion chambers on a series of
10 successful balloon flights during 1995-1999 with a total
exposure of $575 \rm{\,m}^{2} \rm{~hrs}$.  Both experiments were
able to reconstruct proton spectra up to almost 1 PeV.

Figure\,\ref{fig:p-he-fe} shows a collection of the proton,
helium, and iron spectra obtained by various ballon- and
satellite-borne experiments compared to data from ground based
experiments.  Obviously, data from direct experiments are sparse
above 100 TeV and uncertainties become very large with increasing
energies, particularly for primaries heavier than protons.
Reasonably good agreement between RUNJOB and JACEE is observed in
case of the proton spectrum, but the He-flux measured by RUNJOB
is about a factor of two lower compared to other experiments.
Comparing the slopes of the p and He spectra yields power law
indices of $\sim$ 2.7-2.8 for both elements in the energy range
10-500 TeV/nucleon.  The iron spectrum appears somewhat flatter,
$\gamma_{\rm Fe} \simeq 2.6$, particularly when taking into
account the extrapolation to the EAS data.  Such a dependence of
$\gamma$ could be explained by charge dependent effects in the
acceleration or propagation process.  For example, non-linear
models of Fermi acceleration in supernovae remnants predict a
more efficient acceleration for elements with a large $A/Z$
ratio.  However, uncertainties may still be too large to allow
for definite conclusions about differences in the acceleration
and propagation mechanisms of different primaries.  For
illustration, the effect of an assumed (and possibly
underestimated) uncertainty of $\pm 15$\,\% in the energy scale
is shown by the single error bar in each of the panels.

It is remarkable to see direct measurements and EAS data starting
to overlap each other.  Clearly, EAS data below about
$10^{15}$\,eV are dominated by systematic uncertainties while
direct measurements suffer from statistical ones.  With these
caveats kept in mind, the agreement is very good.  The EAS data
of KASCADE \cite{Antoni-05}, also shown in figure
\ref{fig:p-he-fe}, have been reconstructed based on two different
hadronic interaction models employed in the EAS simulations.
Except perhaps for iron, the uncertainties caused by the
interaction model are of similar size or even smaller than
systematic uncertainties between experiments like JACEE and
RUNJOB.
Also shown in figure \ref{fig:p-he-fe} are proton and helium
spectra derived from emulsion chambers and burst detectors
operated within the Tibet II air-shower array
\cite{Amenomori-06a}.  The results are in rough agreement with the
KASCADE data.  For reasons of clarity, only the results based on
simulations with the CORSIKA \cite{CORSIKA} / QGSJet-model
\cite{QGSJET} are included for the Tibet data.  Those obtained
based on Sibyll \cite{SIBYLL} are similar within their error
bars.  There are some important peculiarities of the Tibet
AS$\gamma$ analysis to be pointed out here.  The data are
compared to EAS simulations assuming in one case a heavy
dominated (HD) and in another case a proton dominated (PD)
composition.  In the HD-model a rigidity dependent knee $E_{\rm
k}=Z \times 1.5 \cdot 10^{14}$\,eV is assumed and in the PD-model
all mass components are assumed to break off at $E_{\rm k}= 1.5
\cdot 10^{14}$\,eV. These assumptions are surprising, since no
experiment ever has observed at break in the spectrum at such low
energies.  Furthermore, the experimental data of Ref.\
\cite{Amenomori-06a} start only at energies above $E \simeq
4\cdot10^{14}$\,eV, i.e.\ significantly above the assumed knee
position.  Moreover, because of insufficient separation power
between proton and helium primaries, the authors have deduced the
proton spectrum first by using a neural network algorithm.  Next,
the proton + helium spectrum has been reconstructed and, finally,
the helium spectrum has been obtained by subtracting the number
of proton events obtained in the first task from the proton +
helium dataset obtained in the second task.  Clearly, there are
huge correlated errors to be expected in the helium spectrum
deduced that way.  Also, it is not clear how the results depend
on the ad-hoc assumptions made about the knee position.  Because
of the steeply reconstructed proton and helium spectrum, the
authors then conclude, that the main component responsible for
the change of the power index of the all-particle spectrum around
$3 \cdot 10^{15}$\,eV is composed of heavy primaries.  However,
there is no proof to this statement as the experiment is almost
blind to heavy particles (detection efficiency of iron $\approx
4$\,\%) .

To conclude this topic, despite some controversy a reasonably
good agreement between direct and EAS experiments has been
achieved in recent years.  At present, EAS experiments at their
threshold energies are limited purely by systematic
uncertainties, while direct measurements suffer mostly from
lacking statistics but also from systematic uncertainties in
determining the absolute energy scale.  There is some hope that
new EAS experiments located at very high altitude will be able to
push the measurements down to lower energies and at the same time
also reduce their systematic uncertainties.  Direct experiments,
on the other hand, may be able to increase their exposure at high
energies.  However, given the very steeply falling spectrum, it
appears unlikely that balloon experiments will be able to extend
the range of measurements beyond 1 PeV any time in the near
future.  Thus, the chance of detecting the knee with direct
measurements of protons to iron on balloons is not likely to
occur without significant increases in the payload and flight
duration capabilities of high altitude balloons.  Even with 50
times the present JACEE p-He exposure one would still be unable
to make definitive measurements about a break in the energy
spectrum beyond 200 - 300 TeV \cite{Cherry-05}.

\subsection{Air shower data at the knee}

As can be seen from figure \ref{fig:all-particle}, a wealth of
data at energies around the knee has been accumulated by a large
number of experiments operated over many years.  It is clearly
noticeable that the data fall into two groups differing by their
fluxes mostly: CASA-MIA, CASA-BLANCA, and DICE (all operated at
Dugway, Utah) show distinctly lower fluxes than Tibet, HEGRA,
EAS-TOP, and KASCADE and Tunka \cite{Budnev-05} (not shown in
figure 1).  This problem has already been addressed in
\cite{Swordy-02} but is still not fully understood.  It may be
related to different observation techniques (charged particles
combined either with muons or with Cherenkov light), differences
in the details of EAS simulations, or to other reasons.  On the
other hand, it should be pointed out that the differences almost
vanish, if one of the groups is shifted by about 15\,\% in their
absolute energy scale, i.e.\ by an amount well within the
systematic uncertainties of the experiments.  The knee energy is
found in all experiments at approximately 3 PeV with the index
changing from $\gamma_{1} = 2.7$ to $\gamma_{2} = 3.1$.  Only
Akeno data are different showing different spectral shapes and a
very sharp knee at $\sim 5$\,PeV.

\begin{figure}[t]
\centerline{\epsfxsize=\columnwidth\epsfbox{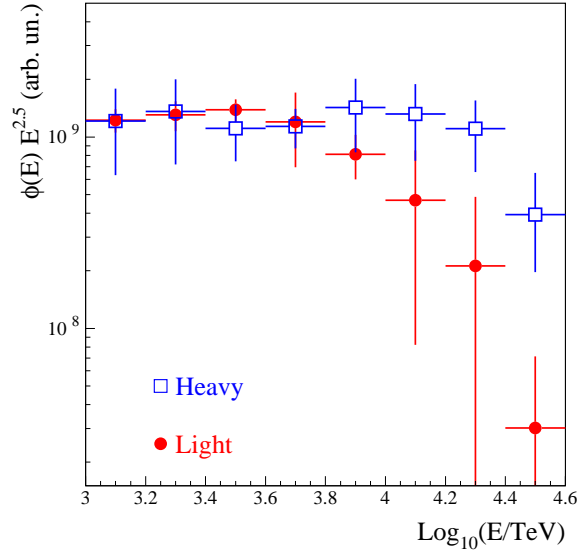}}
\vspace*{-7mm} \caption[xx]{CR energy spectrum of light (p+He) 
and heavy (all the rest) primaries from EAS-TOP and MACRO using 
TeV muons \cite{Aglietta-04}).
\label{fig:e-spec-eas-top}}
\end{figure}

It has been realized that the all-particle spectrum is not very
discriminative against astrophysical models of the knee and that
a deconvolution into different primary particles is required.
However, this is probably the most difficult task in EAS physics,
both because of the level of dependence on hadronic interaction
models used in EAS simulations and because of the significant
(mass dependent) fluctuations of EAS observables.  A large
variety of methods is used to infer the primary energy and mass
\cite{Kampert-01}, most notably the ratio of electron to muon
numbers.  At energies higher than approx.\ $10^{17}$\,eV, also
direct measurements of the shower maximum in the atmosphere
become available by observations of fluorescence light with
imaging telescopes, such as operated by HiRes and the Pierre
Auger Collaboration (see proceedings to this conference).

Extensive analyses of both the energy spectrum and composition
have been performed by EAS-TOP and KASCADE. EAS-TOP has analysed
its data through simultaneous measurements of the electromagnetic
and muonic shower components.  These are obtained from the EAS
array operated at Campo Imperatore on the mountain top 2005
m.a.s.l. (820 g/cm$^{2}$) above the underground Gran Sasso
Laboratories in which the MACRO detector has been located under
an average depth of 1200 m rock \cite{Aglietta-04}.  The
coincident observation of the soft charged particles in the
surface array and the high energy EAS muons ($E_{\mu} >
1.3$\,TeV) in the underground detector permits -- despite large
fluctuations of the muon number -- a reconstruction of the CR
energy spectrum for ``light'' and ``heavy'' primaries.  The
result, depicted in figure \ref{fig:e-spec-eas-top}, shows that 
the energy spectrum of the light primaries is beginning to 
diminish at about 5 PeV, whilst the heavy component may be 
signaling its change in the spectrum at least a decade higher in 
energy.

\begin{figure}[t]
\centerline{\epsfxsize=\columnwidth\epsfbox{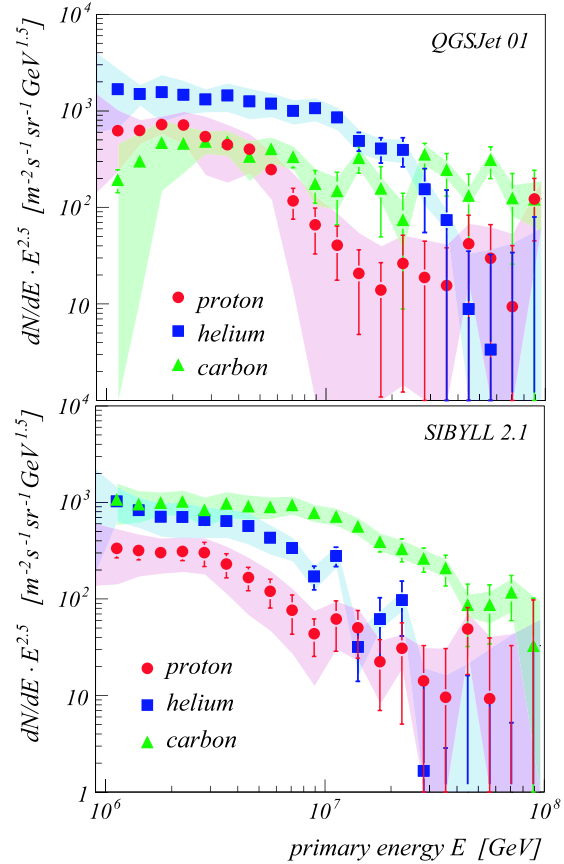}}
\vspace*{-7mm} \caption[xx]{Unfolded CR energy spectrum of p, He,
and C mass-groups from KASCADE. The spectra are obtained by using
QGSJET and SYBILL for the generation of the EAS response matrix
$p_{A}$ \cite{Antoni-05}.
\label{fig:kascade-spectra}}
\end{figure}

The results corroborate those of KASCADE shown in figure
\ref{fig:kascade-spectra}.  KASCADE is located at sea-level (110
m.a.s.l.) in Karlsruhe, Germany, and measures the
electromagnetic, muonic, and hadronic EAS components using a very
dense detector array and a hadronic calorimeter
\cite{KASCADE-NIM}.  The analysis of the data takes advantage of
the effect that for given energy, primary Fe-nuclei result in
more muons and fewer electrons at ground as compared to proton
primaries.  Specifically, in the energy range and at the
atmospheric depth of KASCADE, a Fe-primary yields about 30\,\%
more muons and almost a factor of two fewer electrons as compared
to a proton primary.  The basic quantitative procedure of KASCADE
for obtaining the energy and mass of the CRs is a technique of
unfolding the observed two-dimensional electron vs truncated muon
number spectrum into the energy spectra of primary mass
groups~\cite{Antoni-05}.  The problem can be considered a system
of coupled Fredholm integral equations of the form
\begin{eqnarray*}
\lefteqn{\frac{dJ}{d\,\lg N_e \;\; d\,\lg N_\mu^{\rm tr}} =
\sum_A \int\limits_{-\infty}^{+\infty} \frac{d\,J_A}{d\,\lg E} 
\quad \cdot} \\
& &
\cdot \quad p_A(\lg N_e\, , \,\lg N_\mu^{\rm tr}\, \mid \, \lg E)
  \cdot d\, \lg E
\end{eqnarray*}

\noindent where the probability
\begin{eqnarray*}
\lefteqn{p_A(\lg N_e , \lg N_\mu^{\rm tr}\, \mid \, \lg 
    E) =} \\
& & \int\limits_{-\infty}^{+\infty} k_A(\lg N_e^t , \lg 
    N_\mu^{\rm tr,t})
  d\, \lg N_e^t\,\, d\,\lg N_\mu^{\rm tr,t}
\end{eqnarray*}
is another integral equation with the kernel function $k_A = r_A
\cdot \epsilon_A \cdot s_A$ factorizing into three parts.  Here,
$r_A$ describes the shower fluctuations, i.e.\ the 2-dim
distribution of electron and truncated muon number for fixed
primary energy and mass, $\epsilon_A$ describes the trigger
efficiency of the experiment, and $s_A$ the reconstruction
probabilities, i.e.\ the distribution of $N_e$ and $N_\mu^{\rm
tr}$ that is reconstructed for given {\em true} numbers $N_e^t$,
$N_\mu^{\rm tr,t}$ of electron and truncated muon numbers.  The
probabilities $p_A$ are obtained from CORSIKA simulations using
QGSJET-01 \cite{QGSJET} and Sibyll 2.1 \cite{SIBYLL} as
high-energy and GHEISHA \cite{GHEISHA} as low-energy hadronic
interaction models and a moderate thinning procedure.  Smaller
samples of fully simulated showers were generated for comparison.
The simulated data are then fed into the detector Monte Carlo
programme and the response is parameterized as a function of
energy and mass.  Because of the large shower fluctuations,
unfolding of all 26 energy spectra ranging from protons to
Fe-nuclei is clearly impossible.  Therefore, 5 elements (p, He,
C, Si, Fe) were chosen as representatives for the entire
distribution.  More mass groups do not improve the
$\chi^2$-uncertainties of the unfolding but may result in mutual
systematic biases of the reconstructed spectra.

The results of such an unfolding are presented in figure
\ref{fig:kascade-spectra}.  Shown are the spectra of the p, He,
and C mass-groups based on the response matrices $p_{A}$ obtained
from the two interaction models.  Clearly, there are common
features but also differences in the energy distributions.  In
each of the distributions a distinct break in the spectrum is
observed which is increasing towards higher energy with increasing
primary mass.  In both cases the He flux is higher than the
proton flux.  This finding may be surprising at first sight, but
it is already suggested by extrapolating the He and proton
spectra with their different slopes from lower energies towards
the knee (see figure \ref{fig:p-he-fe}).  The spectrum of the Si
group (see figure 14 and 15 in Ref.\,\cite{Antoni-05}) indicates
a knee at even higher energies.  The Fe spectrum (figure
\ref{fig:p-he-fe}) shows large differences when performing the
unfolding either with the QGSJET or Sibyll model demonstrating
that such analyses are limited at present mostly by uncertainties
of the hadronic interaction models.  Despite these differences in the
individual spectra, the all-particle spectra of KASCADE (see
figure \ref{fig:all-particle}), obtained by summing up the energy
spectra of all mass groups (p - Fe) coincide very nicely for the
two interaction models.  Thus, it can firmly be stated that the
knee in the all particle spectrum is caused by light (p and He)
primaries.  Obviously, also the mean mass composition (e.g.\
expressed in terms on the mean logarithmic mass
\cite{Kampert-02}) increases above the knee.

\begin{figure}[t]
\centerline{\epsfxsize=\columnwidth\epsfbox{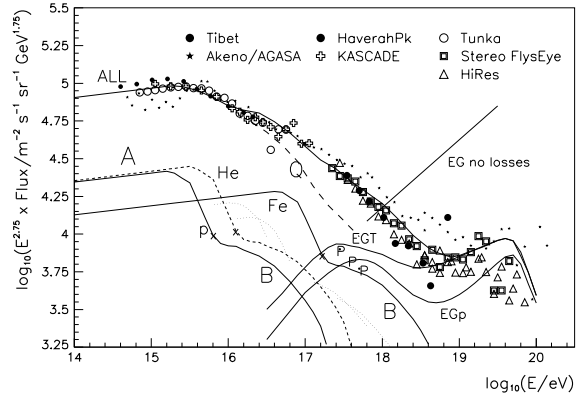}}
\vspace*{-7mm} \caption[xx]{Interpretation of the CR spectrum in
terms of different sources \cite{Hillas-05}.  Shown are the
individual galactic sources (component A and B) and the flux
expected from extragalactic sources.  The galactic components are
guided by the KASCADE knee shape as far as the point marked x.
The dashed line Q is the total if the extended tail B of the
galactic flux is omitted.
\label{fig:Hillas-e-spec}}
\end{figure}

A more detailed investigation \cite{Antoni-05} shows that the
QGSJET model performs reasonably well at high energies but
exhibits some problems at PeV energies.  Sibyll, on the other
hand, describes the data rather well in the knee region but
suffers from a muon deficit at higher energies.  Therefore, it
suggests a more prominent contribution of heavy primaries at high
energies.  It should be emphasized, that this muon deficit of
Sibyll applies to $\mathcal{O}(1 {\rm GeV})$ muons only.  Muons
at energies of several 100 GeV, such as observed by underground
experiments like AMANDA and IceCube, seem to be described rather
well by Sibyll \cite{Montaruli}.  Very recently, a new
interaction model, called EPOS has been released
\cite{Pierog-06}.  Most importantly, it provides a better
description of baryon-antibaryon production at high energies.  A
preliminary analysis shows that the muon number increases more
rapidly with energy than in QGSJET or Sibyll with the muon density
being about 40\,\% higher at $10^{18}$ eV compared to
QGSJET-01 calculations.  It will be interesting to repeat the
unfolding of the CR energy spectra employing this model to verify
whether the present deficiencies of the interaction models will
be resolved.

The unfolded KASCADE energy spectra can directly be compared to
phenomenological calculations of astro- and particle physics
related models or can be used to infer information about the CR
sources.  An example is shown in figure \ref{fig:Hillas-e-spec}
taken from Ref.\,\cite{Hillas-05}.  It is concluded that the data
provide support for the supernova picture of CR origin, i.e.\ the
distinct knee near 3 PeV would be related to emission by the
Ôfree expansionÕ phase of SNRs.  However, a question arises about
how to fill the gap from the iron knee at about $10^{17}$~eV to
the ankle at $\sim 5 \cdot 10^{18}$~eV. These CRs may originate
from SN type II explosions into dense stellar winds where the
interaction generates much stronger magnetic fields.  This may
result in rigidities up to at least $10^{17}$ V (component `B' in
figure \ref{fig:Hillas-e-spec}), especially from a few abnormally
high speed/low mass ejections \cite{Hillas-05}.

A very important question is whether the present data allow to
distinguish a knee of constant rigidity ($E/Z$) from that of
constant energy per nucleon ($E/A$), such as is predicted by
particle physics interpretations of the knee or by the cannonball
model.  Unfortunately, $Z/A$ changes only from $0.5$ in case of
He to $0.46$ for Fe nuclei.  Hence, the question about the
rigidity dependence needs to be answered basically by comparing
the energy spectra of p and He primaries.  Ironically, these are
the two primaries which are most strongly affected by EAS
fluctuations, so that their energy resolutions are deteriorated
substantially.  In fact, overlaying the p and He spectra of
figure \ref{fig:kascade-spectra} using $E/Z$ and $E/A$ abscissas
does not give a clear answer; Sibyll exhibits a slight preference
for charge scaling and QGSJET for mass scaling.  It is hoped,
that the situation will improve somewhat with better models
becoming available.  Improving on the data side seems more
difficult because of two reasons: statistical errors are already
much smaller than systematical ones and (presently not yet
included) data from larger zenith angles are subject to even
stronger EAS fluctuations.

\begin{figure}[t]
\centerline{\epsfxsize=\columnwidth\epsfbox{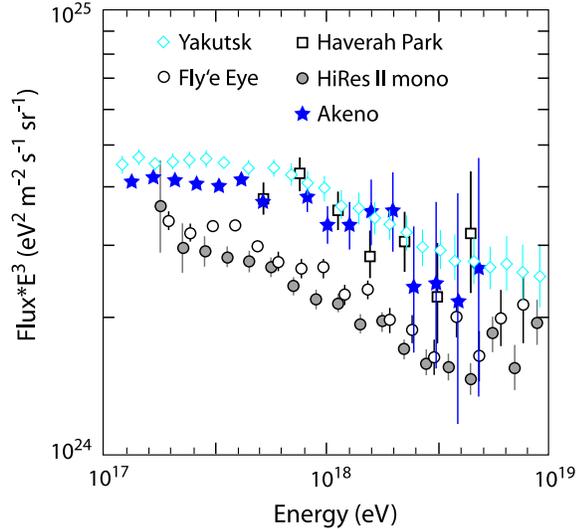}}
\vspace*{-7mm} \caption[xx]{All-particle CR energy spectra from
Yakutsk \cite{Glushkov-05}, Haverah Park \cite{Lawrence-91}, 
Fly's Eye \cite{Bird-93}, HiRes II (mono) \cite{Thomson-06}, and 
Akeno \cite{Nagano-92}.
\label{fig:Fe-knee}}
\end{figure}

\section{THE SECOND KNEE AND ANKLE: TRANSITION FROM GALACTIC TO EXTRAGALACTIC COSMIC RAYS}

Besides the prominent knee in the all-particle spectrum,
additional structures are observed at about $10^{17}$ eV and
$\sim 3 \cdot 10^{18}$ eV, known as the second knee and the
ankle, respectively (see Figs.\ \ref{fig:all-particle} and
\ref{fig:Hillas-e-spec}).  The ankle has been reported
convincingly by a number of experiments, but there is still no
consensus about the existence of a second knee.  This is because
of both the weakness of the structure making it difficult to
detect and because of only few experimental data, most of which
are either at their upper or lower limit of detectable energies.
A blow-up of the data between $10^{17}$ eV and $10^{19}$ eV is
shown in figure \ref{fig:Fe-knee}.  It includes measurements by
Akeno \cite{Nagano-92}, Fly's Eye (stereo) \cite{Bird-93},
Haverah Park \cite{Lawrence-91}, Yakutsk \cite{Glushkov-05}, and
HiRes II (mono) \cite{Thomson-06}.  Akeno has provided the first
hint of a change in the index of the power-law energy spectrum
around $6 \cdot 10^{17}$ eV. The steepening of the spectrum was
confirmed by Haverah Park and is indicated also in the Fly's Eye
and more recent HiRes data.  A recent re-analysis of the Yakutsk
1974-2004 data agrees well with the Akeno data providing
additional support for the existence of a second knee at about
$(6 \pm 2) \cdot 10^{17}$ eV. The ankle at $\sim 3 \cdot 10^{18}$
eV was first observed by Haverah Park, Akeno, and Yakutsk and is
traditionally explained in terms of the transition from galactic
to EGCRs.  The key point here is that one expects the galactic
magnetic field to lose its efficiency at about this energy as the
gyro-radius of a particle at charge $Z$ in a $\mu$-Gauss field,
$r_{g} \simeq 1 {\rm ~kpc} Z^{-1} B_{\mu{\rm G}}^{-1}$, becomes
comparable to the thickness of the galactic disk.  It then
becomes natural to think of hard EGCRs starting to penetrate into
the galaxy and dominating the flux at higher energies (see figure
\ref{fig:Hillas-e-spec} for illustration).

Figure \ref{fig:Hillas-e-spec} also provides an intuitive
explanation for the second knee: it would primarily be caused by
the break of the galactic Fe component.  As Hillas pointed out
\cite{Hillas-05}, an extra component `B' would be needed in order
to make up the well-measured total CR flux at several $10^{17}$
eV for which he considered SNae Type II explosions into dense
stellar winds (see chapter 2.2).  Na\"{i}vely, the second knee in
this picture is expected at $E_{\rm Fe} \simeq 26 \times
3\cdot10^{15} \simeq 8 \cdot 10^{16}$ eV or even lower if the
knee is composed of p and He primaries as suggested by figure
\ref{fig:kascade-spectra}.  This is almost a factor of 10 lower
than reported by Akeno and others.  A scaling of the knee
position with $E/A$ would bring the Fe-knee up to approx.\ $2
\cdot 10^{17}$ eV, but still too low to fit the classical
picture.

Ignoring this puzzle for a moment, also characteristic changes of
the CR composition are expected in this traditional picture of
the knees and ankle.  Up to the knee, the composition would
follow the standard source composition dominated by p and He
primaries.  Between the first and second knee the composition
would change to become iron dominated, and above the ankle it
would be dominated by extragalactic protons.

However, the `folklore' about the second knee and ankle and its
related transition from galactic to EGCRs is not free of dispute
and has received much attention recently.  Back in the 80s,
Berezinsky and collaborators have pointed out an inevitable
feature of the $10^{18}$-$10^{19}$ eV EGCR spectrum: if EGCRs
consist of protons mostly, they would suffer - besides the GZK
effect - from energy losses associated with the production of
$e^{+}e^{-}$ pairs in the CMB photon field \cite{Berezinsky-88}.
This would result in a modulation of the all-particle energy
spectrum to what is called a ``pair-production dip'' between $1
\cdot 10^{18}$ - $4 \cdot 10^{19}$ eV. In such a way, the turn
over from the left- to the right hand side of the `dip' would
mimic the ankle.  Moreover, since the Bethe-Heitler pair
production works effectively only for protons \cite{Aloisio-06},
the ankle can then be interpreted as a signature of a pure proton
EGCR component and the galactic-extragalactic transition must
occur at much lower energies than in the traditional picture,
possibly around the second knee.

\begin{figure}[t]
\centerline{\epsfxsize=\columnwidth\epsfbox{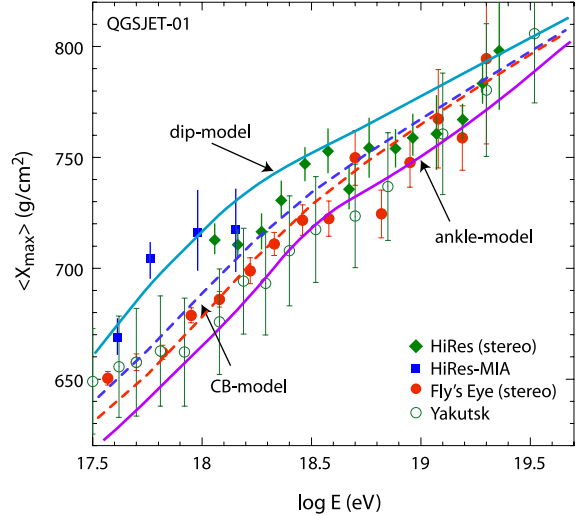}}
\vspace*{-7mm} \caption[xx]{Comparison of the mean depth of
shower maximum, $X_{\rm max}$, predicted by the dip- and
ankle-model of Ref.\ \cite{Allard-06} and the Cannonball of Ref.\
\cite{Dar-06} with data from HiRes \cite{Abbasi-05},
\cite{Abu-Z-00}, Fly's Eye \cite{Bird-93}, and Yakutsk
\cite{Afanasiev-93} (see text for details).
\label{fig:Xmax-comparison}}
\end{figure}

How can the two models be discriminated?  The most critical
observation is provided by a measurement of the chemical
composition in the energy range around $10^{18}$ eV: In the dip
model a strong dominance of protons, and in the ankle model a
strong dominance of iron nuclei is expected.  A recent
confrontation of the two models to existing data has been
performed by Allard \etal~\cite{Allard-06}.  The authors conclude
that the all-particle energy spectrum is reproduced equally well
by the two models.  However, based on a comparison of the mean
mass composition, analyzed in terms of the mean depth of the
shower maximum, $X_{\rm max}$, they favour the traditional model.
Figure \ref{fig:Xmax-comparison} compares the $X_{\rm max}$ data
of various experiments with the dip- and ankle-model of
\cite{Allard-06}.  Here, only CORSIKA / QGSJET-01 simulations are
shown, because QGSJET-01 is the interaction model providing the
most consistent description of experimental data in this energy
region.  Clearly, this direct comparison with QGSJET-01 does {\em
not\/} seem to give preference to any of the two models.  Also
shown are predictions of the cannonball model for two choices of
penetrability of EGCRs into the Galaxy \cite{Dar-06}.  It should
be noted, that in the latter case, $X_{\rm max}(E)$ is
constructed by a simplified model described in Ref.\
\cite{Wigmans-03} instead of using full EAS simulations.
Evidently, better data are required before definite conclusions
can be drawn about the transition from Galactic to extra-galactic
CRs.

\section{ANISOTROPIES}

Another key observation in cosmic ray astrophysics is the
directional distribution of the particles.  That distribution
will depend on any galactic magnetic fields and hence will be
energy (rigidity) dependent.  However, with very limited
exceptions, which are not individually statistically significant,
there is no observed deviation from isotropy above the knee of
the energy spectrum, and any anisotropies at lower energies are
themselves very small \cite{Clay-97,Antoni-04}.  Probably, the
most comprehensive data at energies from a few to several hundred
TeV have been obtained by the Tibet AS$\gamma$ experiment.
Besides revealing fine details of known anisotropies, the data
support the picture of corotation of low energy CRs with the
local Galactic magnetic environment and they may indicate an
anisotropy around the Cygnus region \cite{Amenomori-06b}. 
However, a contamination of TeV $\gamma$'s in the data sample 
cannot be excluded at present.

A non-uniform distribution in of the arrival directions,
suggestive of a source direction, in the energy range $10^{18.0}$
- $10^{18.4}$ eV has been reported by the AGASA
\cite{Hayashida-99} and similarly by the SUGAR collaboration
\cite{Bellido-01}.  However, neither of those observations on
their own is clearly statistically significant.  Moreover, the
Pierre Auger Collaboration has also started to analyse the
galactic centre region.  These results, obtained with much larger
exposure than of AGASA and SUGAR, do not support that finding and
instead provide an upper bound on a point-like flux of CRs from
the Galactic Centre.  Even in absence of CR point sources, such
data may be regarded as the possible beginning of a new era in
cosmic ray astrophysics in which we can begin directional cosmic
ray astronomy.  The possibility of having a source to observe may
indeed open up new frontiers for the Pierre Auger Observatory
\cite{Antoine,Cuoco-06}.

\begin{figure}[t]
\centerline{\epsfxsize=\columnwidth\epsfbox{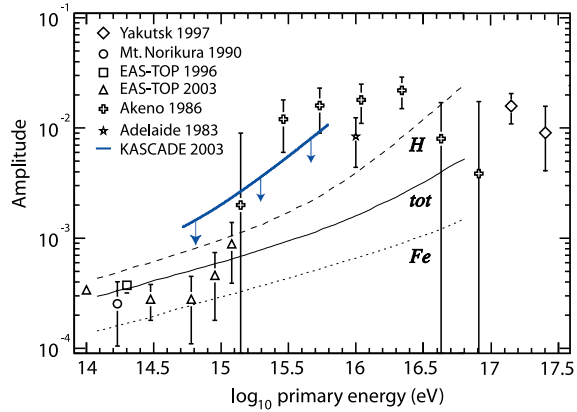}}
\vspace*{-7mm} \caption[]{Rayleigh amplitudes $A$ vs primary
energy from different experiments.  The data of KASCADE
\cite{Antoni-04} (bold line) represent upper limits (95\,\%).
The thin lines show expectations from the galactic CR diffusion
model of Ref.\ \cite{Candia-03}.
\label{fig:Anisotropies}}
\end{figure}

As already pointed out, the low level of CR anisotropy even at
energies above the knee is considered the most serious challenge
to the standard model of the origin of galactic CRs from diffuse
shock acceleration \cite{Hillas-05}.  Figure
\ref{fig:Anisotropies} shows a collection of data expressed in
terms of the Rayleigh amplitudes $A$.  The thin lines represent a
CR diffusion model \cite{Candia-03} predicting anisotropies on a
scale of $10^{-4}$ to $10^{-2}$ depending on particle energy and
strength and structure of the galactic magnetic field.  However,
the model fails to describe the all-particle spectrum
considerably.  Assuming a simple rigidity model of $\tau_{\rm
esc}(E) \propto E^{-0.6}$, Hillas estimates anisotropies at a
level of 5\,\%, 16\,\%, and 180\,\% at $1.5 \cdot 10^{14}$ eV,
$10^{15}$ eV, and $1.5 \cdot 10^{17}$ eV, respectively
\cite{Hillas-05}.  In case of a $E^{-1/3}$ scaling, the values
would go down to 0.6\,\%, 1.1\,\%, and 3.7\,\% which is still in
contradiction to the experimental data of figure
\ref{fig:Anisotropies}.

As already mentioned, the CB-model \cite{Dar-06} predicts much
lower levels of anisotropies than models in which CRs diffuse
away from the central realms of the Galaxy, where most SN
explosions take place.  A CB, on the contrary, is considered a
continuous source of CRs along its trajectory from the galactic
disk into the galactic halo.  Along its trajectory, the source
intensity depends on the local and previously traversed ISM
density.  Thus, the source of CRs is very diffuse and the
directional anisotropy of CRs at the EarthÕs location is expected
to be very small and to vary little with energy.

\section{SUMMARY AND OUTLOOK}

Diffusive shock acceleration in SNRs is considered a viable
mechanism for accelerating cosmic rays and it naturally leads to
a power-law spectrum in rigidity.  However, many fundamental
questions related to the assumption of SNRs being the sources of
galactic cosmic rays are still open.  These questions include,
amongst others, the absence of TeV $\gamma$-radiation from a
large fraction of SNRs, the origin of the knee in the cosmic ray
spectrum, the low level of global anisotropies in their arrival
direction, the transition of galactic to extragalactic CRs and
its related question about the existence of a second knee and
about the origin of the ankle.

The deconvolution of the all-particle CR spectrum into energy
spectra of individual mass-groups by current KASCADE data
\cite{Antoni-05} has advanced the field quite a lot.  Such kind
of data contain much more information than the all-particle
spectrum and the mean mass of CRs (expressed mostly by $\langle
X_{\rm max} \rangle$ and $\langle \ln A \rangle$) alone.
However, there remain large uncertainties, which still allow
alternative interpretations.  Most prominently, a definite answer
about an $E/Z$ (rigidity) or $E/A$ scaling of the knee position
cannot be given at present.  However, there is still some room
for improving the data quality and, despite enormous progress
already made, there are also better hadronic interaction models
being developed which are hoped to eliminate the still existing
shortcomings of the present models such as Sibyll 2.1 or
QGSJET01 \cite{Pierog-06}.

At energies above $10^{17}$ eV data become very sparse and we are
far from understanding the transition from galactic to
extragalactic CRs.  Although the ankle in the CR spectrum at
about $5 \cdot 10^{18}$ eV is often interpreted as the signature
of the transition from a steeply falling galactic CR-spectrum to
a slightly harder extragalactic spectrum, alternative
explanations are possible.  Sometimes the second knee at about
$10^{17.5}$ eV is considered as indication for the transition to
extragalactic CRs, but this explanation would require fine-tuning
of the injection spectra of the different galactic and
extragalactic sources.  Two particular models were discussed in
detail, the dip- \cite{Aloisio-06} and the ankle-model
\cite{Hillas-05}.  Current data on $\langle X_{\rm max} \rangle$
do not allow to exclude any of the two models.  The transition
from galactic to extragalactic CRs occurs in the energy region of
the second knee and is distinctly seen only if iron and proton
spectra are measured separately.

In conclusion, the fundamental question about the origin of high
energy CRs below the GZK energy remains far from being answered.
As a consequence, the interest in studying CRs from about
$10^{17}$ to $10^{19}$ eV with high quality state of the art EAS
detectors has grown worldwide and several new experiments are
being prepared or planned for.  These include KASCADE-Grande
(already in operation) \cite{DiPierro-06} as well as low-energy
extensions of Auger by High Elevation Auger Telescopes (HEAT) and
an infill array with extra muon detectors, as well as the
Telescope Array (TA) and its low-energy extension TALE
\cite{Martens-06}.  These detectors can reliably solve the
problem of measuring the energy spectrum and mass composition in
the transition region and complement the measurements performed
at the highest energies by the Pierre Auger Observatory.

\medskip \noindent {\bf Acknowledgments } Its a pleasure to thank
the organizers for their invitation to the CRIS 2006 workshop
which was conducted in a very pleasant and fruitful atmosphere.
The author is grateful to M. Risse for carefully reading the
manuscript.  The work of the group at University Wuppertal is
supported in part by the Helmholtz VIHKOS Institute and by the
German Ministry for Research and Education.

\end{document}